# Bifurcating trajectory of non-diffractive electromagnetic Airy pulse

**Alexander G. Nerukh,[1*] Denis A. Zolotariov,[1] Dmitry A. Nerukh[2]**

[1]*Kharkov National University of Radio Electronics, 14 Lenin Ave., Kharkov, 61166 Ukraine*
[2]*Non-linearity and Complexity Research Group, Aston University, Birmingham, B4 7ET, UK*
[*]*Corresponding author: nerukh@ddan.kharkov.ua*

The explicit expression for spatial-temporal Airy pulse is derived from the Maxwell's equations in paraxial approximation. The trajectory of the pulse in the time-space coordinates is analysed. The existence of a bifurcation point that separates regions with qualitatively different features of the pulse propagation is demonstrated. At this point the velocity of the pulse becomes infinite and the orientation of it changes to the opposite.

Intensive theoretical and experimental investigations of Airy beams are motivated by their unusual features (non-diffractive propagation, accelerating motion, and self-healing). A solution to the Schrodinger equation in the form of the non-spreading accelerating Airy wave function found by Berry and Balazs in 1979 [1] inspired Siviloglou et al to put forward the concept of electromagnetic Airy beams with similar properties [2,3]. This idea is based on the analogy of the paraxial equation describing the optical beams with the time dependent Schrodinger equation if the time variable is replaced with the spatial coordinate along which the beam propagates. These seminal publications with theoretical formulation and experimental confirmation were followed by many works on the Airy beam properties, for example [4-11] (and citations therein). These investigations generate very interesting applications, some of which are already realised [12-19].

Importantly, the solution to the Schrodinger equation in [1] gives a time dependent Airy wave packet in vacuum with parabolic connection between the spatial and the temporal variables. However, the further analysis of the mathematical aspects and the physical interpretation of the electromagnetic Airy beams considers the *harmonic* time dependence, $\sim \exp(i\omega t)$, while the parabolic connection is considered between the *spatial* coordinates. Only a few papers consider a more general temporal dependence [20-25], and even in the general approach developed in [21] the time dependence corresponds to a Fourier spectrum near some central frequency.

A time dependent problem for an anomalously dispersive medium has been considered in [22-25]. There the only time dependent term in the paraxial wave equation contains the dispersion coefficient which turns to zero if there is no dispersion. The equation becomes time independent in this case. The provided solution, however, is explicitly time dependent even when there is no dispersion. This means that there is an apparent disagreement between the equation and the solution.

Thus, a direct derivation of an explicitly time dependent spatial-temporal Airy pulse in vacuum is needed. The expression must be derived directly from the Maxwell equations and must not use any analogies.

The provocative solution to the Schrodinger equation

$$i\hbar \frac{\partial}{\partial t}\Psi(z,t) + \frac{\hbar^2}{2m}\frac{\partial^2}{\partial z^2}\Psi(z,t) = 0 \qquad (1)$$

contains the Airy function $\mathrm{Ai}[x]$

$$\psi(z,t) = \mathrm{Ai}\left[\frac{B}{\hbar^{2/3}}\left(z - \frac{B^3 t^2}{4m^2}\right)\right] \exp\left[i\frac{B^3}{2m\hbar}\left(tz - \frac{B^3 t^3}{6m^2}\right)\right]. \quad (2)$$

This wave function describes the wave packet propagating in vacuum along the z axis. It is non-diffracting and self-accelerating, which follows from the argument of the Airy function.

The analogy between the Schrodinger equation and the paraxial wave equation describing the two dimensional problem for a monochromatic electromagnetic wave

$$i\frac{\partial}{\partial \xi}\phi(\xi,s) + \frac{1}{2}\frac{\partial^2}{\partial s^2}\phi(\xi,s) = 0 \qquad (3)$$

has been noticed by Siviloglou et al [2, 3]. This analogy is evident if the longitudinal coordinate $\xi$ in (3) plays the role of time $t$ in (1) and the transversal spatial coordinate $s$ in (3) plays the role of the longitudinal coordinate $z$ in (1).

The solution to (3) is a function analogous to the solution (2). It describes an optical beam with a parabolic form in space but a harmonic change in time

$$\phi(\xi,s) = \mathrm{Ai}\left[s - \left(\frac{\xi}{2}\right)^2\right] \exp\left\{\frac{i\xi}{2}\left[s - \frac{\xi^2}{6}\right]\right\}. \qquad (4)$$

To obtain a general explicit time dependent solution of the electromagnetic problem in the form of an Airy pulse we start with the wave equation, followed directly from the Maxwell equation,

$$\partial_{zz}^2 E(t,z) - c^{-2}\partial_{tt}^2 E(t,z) = 0, \qquad (5)$$

which describes the electric field of a wave propagating along the $z$ axis. We will find the solution to this equation assuming the arbitrary time dependence $E(t,z) = F(t,z)e^{ikz}$ with $k$ as the spatial parameter of the wave. In the paraxial approximation $|F_{zz}''| \ll |2ikF_z'|$ [26] the wave equation (5) is reduced directly to the form

$$-i2\partial_\xi F(\tau,\xi) + \partial_{\tau\tau}^2 F(\tau,\xi) + \kappa^2 F(\tau,\xi) = 0, \qquad (6)$$

where the normalised dimensionless variables are $\xi = z/(kc^2 t_0^2)$ and $\tau = t/t_0$ with $t_0$ being the temporal scale. Comparing with (1), the roles of the time and space variables in the electromagnetic time paraxial equation (6) are opposite to those in the Schrodinger equation.

The equation (6) contains also the term with the dimensionless parameter $\kappa = kct_0$ that is similar to the Schrodinger equation with a uniform potential. This term can be eliminated by a proper choice of the factor to the sought solution $F(\tau,\xi)$. However, our numerical investigations show that this parameter has a significant effect on the feasibility of the paraxial approximation $|F_{\xi\xi}''| \ll |2\kappa^2 F_\xi'|$. For small values of the parameter there is a substantial region in the time-space coordinates where the approximation is not fulfilled.

We construct the solution to the equation (6) in the form $B = W(\eta)e^{i\Theta(\eta,\xi)}$ following the procedure described in [11]. Here $W(\eta)$ and $\Theta(\eta,\xi)$ are real functions of the argument $\xi$ and the parabolic variable $\eta = -a\tau + \tau_0 - \xi^2/4 + b\xi$. The parameters $a = \pm 1$, $\tau_0$ and $b$ are very important because, as we show below, they describe extraordinary points that define the whole evolution of the solution. After algebraic manipulations we obtain the expression for the phase

$$\Theta(\eta,\xi) = \pm a^{-2}(-\xi/2 + b)\eta \mp$$
$$2^{-1}a^{-2}\left(\xi^3/12 - b\xi^2/2 + (b^2 + a^2\kappa^2)\xi\right) \qquad (7)$$

and the equation for the envelope $W''(\eta) - \eta W(\eta) = 0$, which has the solution as the Airy function

$$W(\eta) = \mathrm{Ai}[-(\xi/2-b)^2 - a\tau + \tau_0 + b^2]. \qquad (8)$$

These expressions form the solution to the equation (5) in the paraxial approximation:

$$E = F(\tau,\xi)e^{i\kappa^2 \xi} = \mathrm{Ai}\left[-(\xi/2-b)^2 - a\tau + \tau_0 + b^2\right] \times$$
$$\exp\left\{i\left[(a\tau - \tau_0 - b^2)(\xi/2-b) + \right.\right. \qquad (9)$$
$$\left.\left. \xi^3/12 - b\xi^2/2 + (2b^2 + \kappa^2)\xi/2 - b^3\right]\right\}$$

The analysis of this field depends on the sign of $a$, the pulse propagation direction, which determines the direction of the parabola and the location of the wave source. Let us assume that the source of radiation is located at the point $2d$ and $d < b$, Fig. 1.

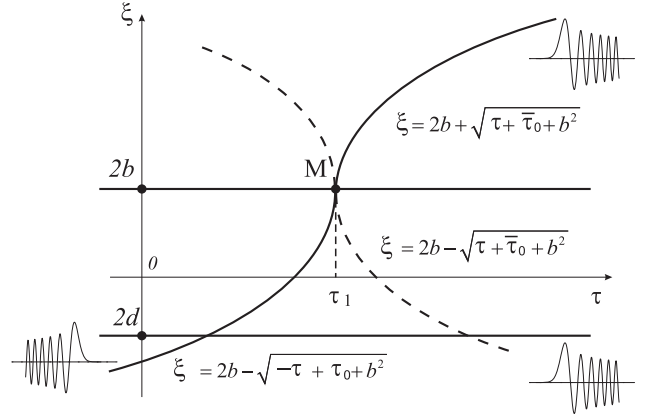

Fig. 1. The trajectory of the pulse and the locations of the source points.

If we consider the propagation in the positive direction of $z$ then we must choose $a = 1$ in (9) because only in this case the propagation corresponds to the positive axis of time $\tau$. The time change of the field in the source

$$E(\tau, \xi = 2d) = \mathrm{Ai}\left[-(d-b)^2 - \tau + \tau_0 + b^2\right] \times$$
$$\exp\left\{i\left[-(\tau + \tau_0 + b^2)(d-b) + \right.\right. \qquad (10)$$
$$\left.\left. d^3/12 - bd^2/2 + (2b^2 + \kappa^2)d - b^3\right]\right\}$$

gives the field radiated by this source along the positive direction of the z axis:

$$E(\tau,\xi) = \mathrm{Ai}\left[-(\xi/2-b)^2 - \tau + \tau_0 + b^2\right] \times$$
$$\exp\left\{i\left[-(\tau + \tau_0 + b^2)(\xi/2-b) + \right.\right. \qquad (11)$$
$$\left.\left. \xi^3/12 - b\xi^2/2 + (2b^2 + \kappa^2)\xi/2 - b^3\right]\right\}$$

Starting from the source point $\xi_1 = 2d$ the field profile propagates along the bottom branch of the left parabola $-\tau + \tau_0 - (\xi/2-b)^2 + b^2 = const$, Fig. 1. The sign of the parabolic variable $\eta$ determines the Airy pulse orientation such that when $\eta$ is positive inside the parabola it moves with the main lobe of the envelope ahead and the oscillating tail at the rear. The velocity of this movement increases with distance $\dot{\xi} = -2/(\xi - 2b)$ up to the point $\xi_0 = 2b$ where it turns to infinity.

After reaching the point $\xi_0 = 2b$ at the moment $\tau_1 = 2const + b^2 - \overline{\tau}_0$ the pulse can propagate along any of the two branches of the right parabola $\tau + \overline{\tau}_0 - (\xi/2 - b)^2 + b^2 = const$ preserving the positive flow of time. This parabola corresponds to the parameter $a = -1$. The point $M(\tau_1, 2b)$ in Fig. 1 is a bifurcation point. The field at this point becomes a time varying source located at $\xi_0$:

$$E(\tau, \xi = 2b) = \text{Ai}[\tau + \overline{\tau}_0 + b^2] \exp\{i(b\kappa^2 - b^3/3)\}. \quad (12)$$

The initial moment $\overline{\tau}_0$ is not arbitrary, but should be chosen $\overline{\tau}_0 = 2const - 2b^2 - \tau_0$ which provides the continuity of the field at the bifurcation point directing the pulse propagation along the upper or the lower branches of the right hand side parabola, Fig. 1. The propagation along the upper branch corresponds to the passage of the pulse through the bifurcation point, while the propagation along the lower branch corresponds to the reflection of the pulse at this point. The field on both trajectories is determined by (9) with $a = -1$:

$$\begin{aligned} E = F(\tau, \xi) e^{i\kappa^2 \xi} = &\text{Ai}\left[-(\xi/2 - b)^2 + \tau + \overline{\tau}_0 + b^2\right] \times \\ &\exp\{i[-(\tau + \overline{\tau}_0 + b^2)(\xi/2 - b) + \\ &\xi^3/12 - b\xi^2/2 + (2b^2 + \kappa^2)\xi/2 - b^3]\} \end{aligned} \quad (13)$$

The point $M(\tau_1, 2b)$ also delimits the regions of the pulse orientation. The pulse moves with the main lobe ahead on the bottom branch of the left parabola $\xi = 2b - \sqrt{-\tau + \tau_0 + b^2}$, turns over at the point $M(\tau_1, 2b)$, and moves further with the tail ahead on any of the branches $\xi = 2b \pm \sqrt{\tau + \overline{\tau}_0 + b^2}$.

The movement of the pulse after the moment $\tau_1 = 2const + b^2 - \overline{\tau}_0$ is slowing, the acceleration $\ddot{\xi} = -4/(\xi - 2b)^3$ is negative, that leads to its complete stop at the infinite distance from the source as its velocity $\dot{\xi} = 2/(\xi - 2b)$ tends to zero.

Summarising, time dependent electromagnetic Airy pulses that satisfy the 'paraxial' equation similar to the Schrodinger equation in which the time and space variables interchange their roles are obtained. The analysis reveales the peculiarities of the Airy pulse propagation and shows the existence of a bifurcation point on the time-space diagram which separates the regions of qualitatively different character of the pulse propagation.

**References**
[1] M.V. Berry, N.L Balazs, Am. J. Phys., 47, 264 (1979).
[2] G. A. Siviloglou, J. Broky, and D. N. Christodoulides, Opt. Lett., 32, 979 (2007).
[3] G. A. Siviloglou, J. Broky, A. Dogariu, and D. N. Christodoulides, Phys. Rev. Lett., 99, 213901 (2007).
[4] I.M. Besieris, A.M. Shaarawi, Opt. Lett., 32, 2447 (2007).
[5] P. Saari, Opt. Express, 16, 10303-10308 (2008)
[6] L. Carretero, P. Acebal, S. Blaya, C. García, A. Fimia, R. Madrigal, and A. Murciano, Opt. Express, 17, 22432 (2009)
[7] M.A. Bandres, Opt. Lett., 34, 3791 (2009).
[8] S. Vo, K. Fuerschbach, K. P. Thompson, M. A. Alonso, and J. P. Rolland, J. Opt. Soc. Am. A, 27, 2574 (2010)
[9] Y. Kaganovsky and E. Heyman, Optics Express, 18, 8440 (2010)
[10] M.I. Carvalho, M. Facao, Opt. Express, 18, 21938 (2010).
[11] Chi-Young Hwang, Dawoon Choi, Kyoung-Youm Kim, and Byoungho Lee, Opt. Express, 18, 23504 (2010)
[12] J. Baumgartl, M. Mazilu, K. Dholakia, Nature photonics, 2, 675 (2008).
[13] T. Ellenbogen, N. Voloch-Bloch, A. Ganany-Padowicz, A. Arie, Nature photonics, 3, 395 (2009).
[14] A. Salandrino and D. N. Christodoulides, Optics Letters, 35, 2082 (2010).
[15] A. Salandrino and D. N. Christodoulides, Optics Letters, 36, 487-489 (2011).
[16] H. Cheng, W. Zang, W. Zhou, and J. Tian, Opt. Express, 18, 20384 (2010)
[17] J.-X. Li, W.-P. Zang, and J.-G. Tian, Opt. Express, 18, 7300-7306 (2010)
[18] Z. Zheng, B.-F. Zhang, H. Chen, J. Ding, and H.-T. Wang, Apll. Optics, 50, 43-49 (2011).
[19] Wei Liu, D. N. Neshev, I. V. Shadrivov, A. E. Miroshnichenko, and Yu. S. Kivshar, Opt. Lett., 36, 1164-1166 (2011)
[20] D.N. Christodoulides, N.K. Efremidis, P.D. Trapani, B.A. Malomed, Opt. Lett., 29, 1446 (2004).
[21] I. M. Besieris, A. M. Shaarawi, Phys. Rev. E, 78, 046605(1-6) (2008).
[22] A. Chong, W. H. Renninger, D. N. Christodoulides and F. W. Wise, Nature photonics, 4, 103 (2010)
[23] T. J. Eichelkraut, G. A. Siviloglou, I. M. Besieris, and D. N. Christodoulides, Opt. Letters, 35, 3655 (2010).
[24] D. Abdollahpour, S. Suntsov, D. G. Papazoglou, and S. Tzortzakis, Phys. Rev. Letters, 105, 253901-4 (2010).
[25] D. G. Papazoglou, N. K. Efremidis, D. N. Christodoulides, and S. Tzortzakis, Opt. Lett., 36, 1842 (2011).
[26] Y.B. Band, Light and matter, John Willey & Sons, Inc. (2007).